\documentclass[aps,prb,twocolumn,floatfix,footinbib]{revtex4}
\usepackage{epsfig}
\usepackage{amsfonts}
\usepackage{amsmath}
\usepackage{amssymb}
\usepackage{graphicx}%
\begin{document}
\title{Spin Modulation in Semiconductor Lasers}
\author{Jeongsu Lee$,^{1}$, William Falls,$^{1}$ 
Rafal  Oszwa\l dowski,$^{1,2}$ 
and Igor \v{Z}uti\'{c}$^{1}$}
\affiliation{
$^{1}$ Department of Physics, State University of New York 
at Buffalo, NY, 14260, USA \\
$^{2}$ Instytut Fizyki, Uniwersytet Miko{\l }aja Kopernika, Grudzi\c{a}dzka
5/7, 87-100, Toru\'n, Poland}
%\date{\today}

\begin{abstract}
We provide an analytic study of the  dynamics of semiconductor lasers with 
injection (pump) of spin-polarized electrons, previously considered  
in the steady-state regime. 
Using complementary approaches of quasi-static and small signal analyses,
we elucidate how the spin modulation in semiconductor lasers can 
improve performance, as compared to the conventional (spin-unpolarized)
counterparts. We reveal that the spin-polarized injection can 
lead to an enhanced bandwidth and desirable switching properties of
spin-lasers. 

\end{abstract}
\maketitle
Practical paths to spin-controlled devices are typically limited to 
magnetoresistive effects, successfully employed for magnetically
storing and sensing information.\cite{Zutic2004:RMP,Fabian2007:APS} 
However, spin-polarized carriers generated 
in semiconductors by 
circularly polarized light  
or electrical 
injection,\cite{Zutic2004:RMP,Meier:1984} can also enhance the 
performance of lasers, 
for communications and signal 
processing. While such spin-lasers  
already demonstrate 
a lower threshold current,\cite{Rudolph2003:APL,Holub2007:PRL,Hovel2008:APL} 
as compared to their conventional (spin-unpolarized) 
counterparts, many theoretical challenges remain.  Even in the steady-state 
regime, several surprises have only recently been revealed. For example, 
a very short spin relaxation time of holes can be 
advantageous,\cite{Gothgen2008:APL} with the maximum threshold reduction 
larger than what was theoretically thought 
possible.\cite{Gothgen2008:APL,Vurgaftman2008:APL}

Some of the most attractive properties of conventional lasers lie in their
dynamical performance.\cite{Chuang:1995}  
Here we explore novel opportunities offered by 
spin-polarized modulation. We generalize the rate equation (RE) 
approach\cite{Chuang:1995,Dery2004:IEEEJQE} to describe 
spin-lasers\cite{Gothgen2008:APL} with a quantum well (QW), typically
GaAs or (In,Ga)As, used as the active 
region.\cite{Rudolph2003:APL,Holub2007:PRL,Hovel2008:APL}
However, our analytical approach allows considering other materials
for spin-lasers.\cite{Oestreich2005:SM}    
Spin-resolved electron and hole densities are $n_\pm$, $p_\pm$,
where $+(-)$ denotes the spin up (down) component; the total carrier 
densities are $n=n_++n_-$, $p=p_++p_-$.  
For photon 
density we write $S=S^++S^-$, where $+(-)$ is the right (left) 
circularly-polarized component. 
Spin-polarized electrons, 
injected/pumped into the QW can be represented by  
a current\cite{norm} (density) $J=J_++J_-$ and the corresponding
current polarization $P_J=(J_+-J_-)/J$. 
Each of these quantities, $X$, 
can be decomposed into a steady-state $X_0$ and a modulated part 
$\delta X(t)$, $X=X_0+\delta X(t)$.  

\begin{figure}[tbh]
\includegraphics[scale=0.7]{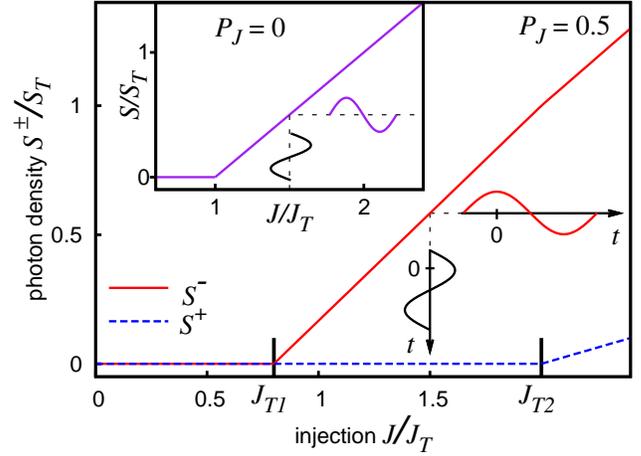}
\caption{(Color online) Amplitude modulation ({\em AM}).
Circularly polarized photon densities, $S^\pm$, as a function of 
electron current $J$ with polarization $P_{J}\equiv P_{J0}=0.5$
(readily realized optically or electrically, using a polarizer), 
for infinite electron spin relaxation time. 
$J$ is normalized to the unpolarized threshold current $J_T$.
Thresholds $J_{T1,2}$ correspond to $S^\mp$ ($J_{T1}<J_T<J_{T2}$).
$S^\pm$ are normalized to the unpolarized photon density $S_T=S(2J_T)$.
Vertical and horizontal harmonic 
curves show the modulation of (input) current and the resulting modulation 
of the (output) light. %, respectively.
{\em AM} of a partially-polarized current $J <J_{T2}$
leads to the modulation of the fully-polarized output light (no $S^+$
component). Inset: 
For {\em AM} in  a spin-unpolarized laser\cite{Chuang:1995} 
($P_J=0$), the photon densities 
$S^+$ and $S^-$ ($S=S^+ + S^-$) undergo identical modulations.} 
\label{fig:AM}
\end{figure}

We focus here on the harmonic amplitude and polarization modulation 
({\em AM}, {\em PM}).
{\em AM} for a steady-state polarization implies $J_+ \neq J_-$ 
(unless $P_J=0$, as in conventional lasers\cite{Chuang:1995}),
\begin{equation}
AM: \:
J=J_0+\delta J \cos(\omega t), \quad  P_J = P_{J0},
\label{eq:AM}
\end{equation}
where $\omega$ is the angular modulation frequency.
{\em AM} is illustrated schematically in Fig.~\ref{fig:AM}. 
Similar to  the steady-state analysis,\cite{Gothgen2008:APL} 
$P_J \neq 0$ leads to unequal threshold currents $J_{T1}$ and $J_{T2}$ 
(for $S^{\mp}$, majority and minority photons). For the injection
$J_{T1}<J<J_{T2}$, we 
expect a modulation of fully polarized 
light, even for a partially polarized injection. 
Such a modulation can be contrasted with {\em PM} 
which also has $J_+ \neq J_-$, but 
$J$ remains constant
\begin{equation}
PM: \:
J=J_0, \quad P_J=P_{J0}+\delta P_J \cos(\omega t).
\label{eq:PM}
\end{equation}
While an experimental 
implementation of the idealized {\em PM} [a fully
time-independent $J$, in Eq.~(\ref{eq:PM})] remains a challenge, 
we analyze it 
theoretically to note its potential advantages.
Just as a decade ago there was 
an early progress towards 
electrical spin injection in semiconductors (now 
well-established\cite{Zutic2004:RMP}), 
a recent progress in electrically  
tunable ${P_J}$\cite{Li2009:APL} is encouraging 
that {\em PM} could
be realized 
in future spin-lasers. Currently,
optically injected lasers with a controllable degree of circular polarization
are more promising for implementing 
{\em PM}.\cite{Rudolph2003:APL,Hovel2008:APL,Ando1998:APL}

In QWs the spin relaxation time\cite{Zutic2004:RMP} 
for holes $\tau_s^p$  is much shorter than for electrons $\tau_s^n$, 
so holes can be considered unpolarized, $p_\pm=p/2$. The charge 
neutrality condition, 
$p_\pm=n/2$,  can then be used to decouple the REs for 
holes from those 
for electrons which become\cite{Gothgen2008:APL}
\begin{eqnarray}
dn_\pm/dt=J_\pm-g_\pm(n_\pm,S)S^\mp
  -\left(n_\pm-n_\mp\right) /\tau_s^n
  -R_{\mathrm{sp}}^\pm, \label{eq:REn} \\
dS^\pm/dt=\Gamma g_\mp(n_\mp,S)S^\pm 
-S^\pm/\tau_{\mathrm{ph}}+\beta\Gamma R_{\mathrm{sp}}^\mp, 
\quad \quad \quad \quad \label{eq:RES}
\end {eqnarray}
where $g_\pm$ is the spin-dependent optical gain, 
$\tau_{\mathrm{ph}}$ is the photon lifetime, $\Gamma$ is the optical
confinement coefficient, $\beta$ is the spontaneous-emission coupling 
coefficient. We consider the linear form of 
radiative spontaneous recombination\cite{Zutic2006:PRL} 
$R_{\mathrm{sp}}^{\pm}=n_\pm/\tau_\mathrm{r}$,
where $\tau_\mathrm{r}$ is the recombination time.\cite{QR} 
In conventional lasers ($P_J=0$), the optical gain term, describing stimulated 
emission, can be 
modeled as\cite{Chuang:1995} $g(n,S)=g_0(n-n_{\mathrm{tran}})/(1+\epsilon S)$, 
where $g_0$ is the density-independent coefficient,\cite{Holub2007:PRL} 
$n_{\mathrm{tran}}$ is the transparency density, 
and $\epsilon$ is the gain compression factor.\cite{Huang1993:OQE}
For $P_J\neq 0$, even 
at $\epsilon=0$, the correct generalization\cite{Gothgen2008:APL} 
$g(n,S) \rightarrow g_\pm(n_\pm,p_\pm,S)=g_0(n_\pm+p_\pm-n_{\mathrm{tran}})$ 
differs from the previous expressions\cite{Rudolph2003:APL,Holub2007:PRL}
but, 
combined with the charge neutrality condition ($p_\pm=n/2$),  
it leads to the correct maximum threshold 
reduction.\cite{Gothgen2008:APL,Vurgaftman2008:APL}  With small experimental 
values of $\epsilon$ in spin-lasers,\cite{Rudolph2003:APL,Holub2007:PRL} 
the gain compression
at moderate pumping intensities is almost negligible. 
Given the typical range\cite{Rudolph2003:APL,Holub2007:PRL} 
of $\beta\sim10^{-5} -10^{-3}$, we mostly focus on the limit $\beta = 0$, 
for which the operating regimes of the spin-lasers can be simply described. 

To develop a preliminary understanding of {\em AM} and {\em PM} in 
spin-lasers, 
we study analytically the quasi-static regime 
($\omega \ll 1/\tau_\mathrm{ph}, 1/\tau_\mathrm{r}$). This implies that 
the steady-state results\cite{Gothgen2008:APL} can be used to 
obtain {\em AM} ({\em PM}) with $J_0$ ($P_{J0}$) substituted by $J(t)$ 
($P_J(t)$)
and the injection $J_\pm$ will be in phase with the 
response $n_\pm$ and $S^\mp$.
For typical parameters,\cite{Holub2007:PRL,Rudolph2003:APL} 
we confirmed numerically  
that this regime is
valid up to $\omega/2 \pi \sim 10$ MHz. 
The steady-state results  ($\epsilon=\beta=0$) 
for the two threshold currents (see Fig.~\ref{fig:AM}), 
\begin{equation}
J_{T1}=J_T/[1+|P_J|/(1+2\mathrm{w})], \quad J_{T2}=J_T/(1-|P_J|),
\label{eq:JT12}
\end{equation}
remain directly applicable for {\em AM} and {\em PM}.
Here $J_{T}=N_{T}/\tau_{\mathrm{r}}$ is the unpolarized threshold current, 
$N_{T}=n\left( J\geq J_{T}\right)  
=\left(  \Gamma g_0 \tau_{\mathrm{ph}}\right)^{-1}+n_{\mathrm{tran}}$ 
is the electron threshold density, and $\mathrm{w}=\tau_\mathrm{r}/\tau_s^n$
is the ratio of the recombination and spin relaxation times.
$J_{T1}$ and $J_{T2}$ in Eq.~(\ref{eq:JT12}) 
delimit three regimes of a spin-laser:   
(i) For $J<J_{T1}$, the laser operates 
as a spin light-emitting diode (LED),\cite{Zutic2004:RMP}
(ii) For $J_{T1}\leq J_T\leq J_{T2}$, there is a mixed operation: lasing
only with one circular polarization 
(only $S^-$, if $P_J>0$), 
while the other circular polarization is still in the spin-LED regime, and 
(iii)  For $J \ge J_{T2}$, the laser is fully lasing with both $S^\pm >0$.

\begin{figure}[tbh]
\includegraphics[scale=0.6]{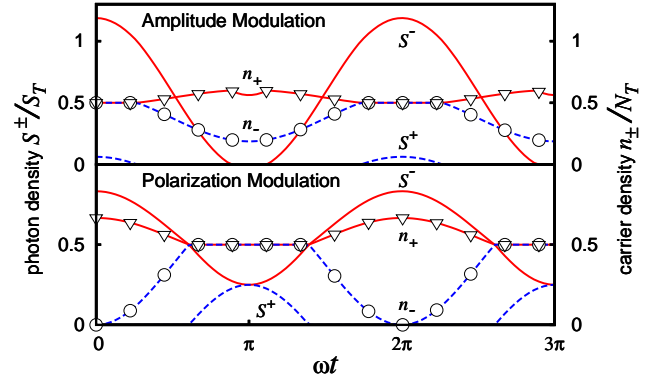}
\caption{(Color online)
Time-dependence  of the normalized photon and carrier densities, calculated 
analytically, in the quasi-static
regime at steady-state current $J_0=1.5 J_T$, 
and polarization $P_{J0}=0.5$. The normalization constants are
$N_T$ and $S_T=J_T\Gamma \tau_{\mathrm{ph}}=S(2 J_T)$. %RO
Large amplitude and polarization modulation is assumed:
$\delta J/J_0=0.5$ and $\delta P_J=0.5$, respectively
[Eqs.~(\ref{eq:AM}) and (\ref{eq:PM})]. 
The results, showing coupling of $n_\pm$ with $S^\mp$
[Eqs.~(\ref{eq:REn}) and (\ref{eq:RES})], %RO
are given for an infinite 
spin relaxation time, $\mathrm{w}=0$, and $\beta=\epsilon=0$.
For $J>J_{T2}$ (a fully lasing regime), $S^+$ is in-phase with $S^-$
for {\em AM} (upper panel) and $S^+$ has an opposite phase to $S^-$ 
for {\em PM} (lower panel), while $n_\pm$ are pinned at $N_T/2$.
}
\label{fig:QSA}
\end{figure}

These operating regimes, for large {\em AM} and {\em PM},
determine the time-dependence
of electron and photon densities,
normalized to  $N_T$ and $S_T=J_T \Gamma \tau_{\mathrm{ph}}$, shown 
in Fig.~\ref{fig:QSA}.
{\em AM} in the upper panel reveals $S^\pm$ near $\omega t=0, 2\pi$, 
corresponding to $J>J_{T2}$ (a fully lasing regime) and a constant 
$n_+=n_-$. With time evolution the laser enters the mixed
regime  $J_{T1}<J<J_{T2}$ (only lasing with $S^-$, Fig.~\ref{fig:AM}).
If $P_J \ge 0$, for both {\em AM} and {\em PM} (discussed below)
the photon densities in Fig.~\ref{fig:QSA} can be expressed as 
\begin{equation}
S^-/S_T=(2/3)\left[J/J_T (1+P_J/2)-1 \right], \: \:  S^+/S_T=0,
\label{eq:mixed}
\end{equation}
where $J$ and $P_J$ are given by either Eq.~(\ref{eq:AM}) or
(\ref{eq:PM}). For $P_J \le 0$, the
expressions for $S^+$ and $S^-$ in Eq.~(\ref{eq:mixed}) are simply exchanged.
Finally, near $\omega t= \pi,3\pi$, no emitted $S^{\pm}$ 
implies $J<J_{T1}$ (the spin-LED regime).
{\em PM} in the lower panel shows only $J>J_{T1}$ 
($S^-$ is present). 
The fully
lasing regime, 
near $\omega t=\pi, 3\pi$,  corresponds to 
constant $n$, even as $P_J(t)$ varies.

The quasi-static approach allowed us to consider analytically large signal
modulation (both for {\em AM} and {\em PM}), usually only studied numerically. 
We next turn to the complementary approach for laser dynamics, i.e., 
small signal analysis (SSA), limited to a small modulation 
($|\delta J/J_0| \ll 1$ 
for AM and $|\delta P_J| \ll 1,  |P_{J0}\pm \delta P_J|<1$ for 
{\em PM}\cite{pure}) 
but valid for all frequencies. 
We confirmed that the two approaches coincide
in the common region of validity. 
Our SSA for spin-lasers proceeds as in conventional 
lasers.\cite{Chuang:1995,Yariv:1997}
The decomposition 
$X=X_0+\delta X(t)$ for $J_\pm$, $n_\pm$, and $S^\pm$ is substituted in 
Eqs.~(\ref{eq:REn}) and (\ref{eq:RES}). The modulation terms 
can be  written as
$\delta X(t)=Re[\delta X(\omega)e^{-i\omega t}]$. 
We  then analytically calculate 
$\delta n_\pm (\omega)$,  $\delta S_\pm(\omega)$ and the appropriate 
generalized frequency response functions 
$R_\pm(\omega)=|\delta S^\mp(\omega)/\delta J_\pm(\omega)|$,
which reduce to the 
conventional form~\cite{Chuang:1995} 
$R(\omega)=|\delta S(\omega)/\delta J(\omega)|$,
in the $P_J=0$ limit.
From SSA we obtain the 
resonance in the modulation response $R_\pm(\omega)$, also
known as ``relaxation oscillation frequency,'' $\omega_R/2 \pi$.
For $P_{J0}=0$ and $J>J_{T2}$, we find
\begin{eqnarray}
\omega^{AM}_R\approx \sqrt{2}\omega^{PM}_R &\approx&
\left[\Gamma g_0 (N_T/\tau_r)(J_0/J_T-1)\right]^{1/2} \nonumber \\
&=&(g_0 S_0/\tau_{ph})^{1/2},
\label{eq:om3}
\end{eqnarray}
where $\omega_R^{AM}$ recovers the standard result.\cite{Chuang:1995}
For 
$P_{J0}\neq0$ and  $J_{T1}<J<J_{T2}$
\begin{eqnarray}
\omega^{AM}_R=\omega^{PM}_R&\approx&
\left[\Gamma g_0 (N_T/\tau_r)(1+|P_{J0}|/2)J_0/J_T-1\right]^{1/2}  \nonumber \\
&=&(3g_0 S_0^-/2\tau_{ph})^{1/2}. 
\label{eq:om2}
\end{eqnarray}
Such $\omega_R$ correspond to a peak in the frequency response
and can be used to estimate its bandwidth,\cite{Chuang:1995} 
a frequency where the normalized response 
$R(\omega/2\pi)/R(0)$ decreases to $-3$ dB.

\begin{figure}[tbh]
\includegraphics[scale=0.7]{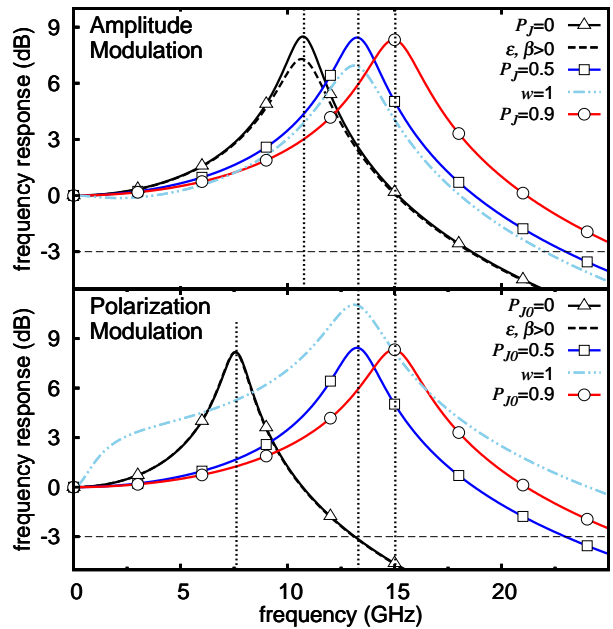}
\caption{(Color online) 
Small signal analysis of amplitude and polarization modulation 
({\em AM}, {\em PM})
for $\delta J/J_0=0.01$ and $\delta P_J=0.01$, at $J_0=1.9 J_T$.
The frequency response function $|\delta S^-(\omega)/J_+(\omega)|$ is
normalized to the corresponding $\omega=0$ value. Results are
shown for $\tau_r=200$ ps, $\tau_\mathrm{ph}=1$ ps, in the  
limit of $\mathrm{w}=0$, and $\beta=\epsilon=0$,
except broken lines for $\mathrm{w}=1$, $\beta=10^{-4}$, and
$\epsilon=2\times 10^{-18}$ cm$^3$. The frequency response at $-3$ dB value
gives the bandwidth of the laser.\cite{Chuang:1995} 
The vertical lines denote approximate
peak positions evaluated at $\mathrm{w}=\epsilon=\beta=0$ from 
Eq.~(\ref{eq:om2}), except
for {\em PM} at $P_{J0}=0$ evaluated from Eq.~(\ref{eq:om3}). When $P_{J0}=0$ 
there is a small difference between $\epsilon=\beta=0$ and finite 
$\epsilon, \beta$ 
results for {\em AM}, but this difference is nearly invisible for {\em PM}. 
} 
\label{fig:SSA}
\end{figure}

We can now look for possible advantages in the dynamic 
operation of spin-lasers, 
as compared to their conventional counterparts.  
From Eq.~(\ref{eq:om2}) we infer that the spin-polarized injection 
increases $\omega_R$ and thus increases the 
laser bandwidth -- an important figure of merit.\cite{Chuang:1995} 
In Fig.~\ref{fig:SSA} 
these trends are visible in the normalized frequency response.
Results for
$P_{J0}=0$ (using finite\cite{Rudolph2003:APL,Holub2007:PRL} 
$\epsilon$ and $\beta$) show that our analytical approximations
for $\epsilon=\beta=0$ are an accurate description of  {\em AM} and {\em PM}, 
at moderate pumping power.
The increase of $\omega_R$ and the bandwidth with $P_{J0}$,  for {\em AM} 
and {\em PM}, can be understood as the dynamic manifestation of threshold 
reduction with increasing $P_{J0}$. 
With $\omega_R \propto (S_0^-)^{1/2}$ [Eq.~(\ref{eq:om2})],
the situation is analogous to the 
conventional lasers: $\omega_R$ and the bandwidth both increase with 
the square root of the output power\cite{Chuang:1995,Yariv:1997} 
($S_0^+=0$ for $J_{T1}<J<J_{T2}$).
An important advantage of spin-lasers is that 
the increase in $S_0^-$ can be achieved even at 
{\em constant} input power (i.e., $J-J_T$), simply by increasing $P_{J0}$.
Additionally, a larger $P_{J0}$ allows for a larger  $J_0$ (maintaining
$J_{T1}<J_0<J_{T2}$), 
which can further enhance the bandwidth, as seen in Eq.~(\ref{eq:om2}).
For $P_{J0}=0.9$, $J_0$ can be up to $10$ $J_T$, from Eq.~(\ref{eq:JT12}). 

We next examine the effects of finite $\tau_s^n$, shown for $\mathrm{w}=1$
and $P_{J0}=0.5$.  {\em AM} results follow a plausible trend: 
$\omega_R$ and the bandwidth monotonically decrease and eventually attain 
``conventional'' values for $\tau_s^n \rightarrow 0$  
($\mathrm{w} \rightarrow \infty$). 
The situation is rather different for {\em PM}: seemingly detrimental spin 
relaxation {\em enhances} the bandwidth and the peak in the frequency response, 
as compared to the long $\tau_s^n$ limit ($\mathrm{w}=0$). 
A shorter $\tau_s^n$ will reduce $P_J$ and thus the amplitude of 
modulated light. 
Since $\delta S^-(0)$ decreases faster with $\mathrm{w}$
than $\delta S^-(\omega>0)$, we find 
an increase in the normalized response function, shown 
in Fig.~\ref{fig:SSA}. 
The increase in the bandwidth comes at the cost of a reduced modulation signal. 

The above considered trends 
allow us to infer some other possible advantages of {\em PM} at fixed
injection.  In the quasi-static approximation, 
for $J>J_{T2}$,  constant $n_+=n_-=N_T/2$ (Fig.~\ref{fig:QSA}) 
implies that {\em PM} 
would be feasible at a reduced chirp ($\alpha$-factor), since  
$\delta n(t)$, which is a chirp source in {\em AM}, 
is eliminated.\cite{Yariv:1997}
These findings in the limit of low-frequency and 
$\epsilon=\beta=0$, 
can be combined with SSA in Fig.~\ref{fig:SSA}, revealing   
substantially smaller 
$\epsilon$, $\beta$ effects
for {\em PM} than {\em AM}, to  suggest that the reduced chirp
and therefore desirable switching properties of spin-lasers
can be expected for a broad range of parameters.

In this work we predict an improved performance of 
spin-lasers and show how it can be understood from the threshold 
reduction experimentally demonstrated for the steady-state 
regime.\cite{Rudolph2003:APL,Holub2007:PRL,Hovel2008:APL} 
Future advances in spin-lasers may depend on progress 
in magnetic memories and data storage. Answers to the key questions in 
these areas, 
about ultra-fast magnetization dynamics and timescales for 
magnetization reversal,\cite{Garzon2008:PRB} may also determine the switching 
speed limit in the modulation of electrically pumped spin-lasers. 

This work was supported by the U.S. ONR, AFOSR-DCT, and NSF-ECCS CAREER. 
We thank H. Dery, C. G\o thgen, A. Petrou, G. Strasser, and X. Zhang 
for stimulating discussions.

%\linespread{1.8}

%\newpage


\begin{thebibliography}{99}
%\newpage
\bibitem{Zutic2004:RMP}
I. \v{Z}uti\'c, J. Fabian, and S. Das Sarma,  Rev. Mod. Phys. {\bf 76}, 
323 (2004).

\bibitem{Fabian2007:APS}
J. Fabian, A. Mathos-Abiague, C. Ertler, P. Stano, and 
I. \v{Z}uti\'c,  Acta Phys. Slov. {\bf 57}, 565 (2007).

\bibitem{Meier:1984}
{\it Optical Orientation,} edited by F. Meier and B.~P. Zakharchenya
(North-Holland, New York, 1984);
I. \v{Z}uti\'c, J. Fabian, and S. Das Sarma, Phys. Rev. B {\bf 64}, 
121201(R) (2001).

\bibitem{Rudolph2003:APL}
J. Rudolph, D.~H\"{a}gele, H.~M. Gibbs, G. Khitrova, 
and M. Oestreich,  Appl. Phys. Lett. {\bf 82}, 4516 (2003);
J. Rudolph, S. D\"{o}hrmann, D.~H\"{a}gele, M. Oestreich, and W. Stolz,
Appl. Phys. Lett. {\bf 87}, 241117 (2005).

\bibitem{Holub2007:PRL}
M. Holub, J. Shin, D. Saha, and P. Bhattacharya, 
Phys. Rev. Lett. {\bf 98}, 146603 (2007);
M. Holub and P. Bhattacharya, J. Phys. D: Appl. Phys. {\bf 40}, R179 (2007).

\bibitem{Hovel2008:APL}
S. H\"{o}vel, A. Bischoff, N.~C Gerhardt, M.~R. Hofmann, T. Ackemann,
A. Kroner, and R. Michalzik, Appl. Phys. Lett. {\bf 92}, 041118 (2008).

\bibitem{Gothgen2008:APL}
C. G\o thgen, R.  Oszwa\l dowski, A. Petrou, and  
I. \v{Z}uti\'c, Appl.  Phys.  Lett. {\bf 93}, 042513 (2008). 

\bibitem{Vurgaftman2008:APL}
I. Vurgaftman, M. Holub, B.~T. Jonker, and J.~R. Mayer,   
Appl.  Phys.  Lett. {\bf 93} 031102 (2008). 

\bibitem{Chuang:1995}
S.~L. Chuang, {\it Physics of Optoelectronic Devices} (Wiley, New York, 1995).

\bibitem{Dery2004:IEEEJQE}
See, e.g., H. Dery and G. Eisenstein, IEEE J. Quant. Electron. {\bf 40}, 
1398 (2004); M. San Miguel, O. Feng, and J.~V. Moloney, Phys. Rev. A
{\bf 82}, 1728 (1995). 

\bibitem{Oestreich2005:SM}
M. Oestreich, J. Rudolph, R. Winkler, and D. H\"{a}gele, Superlattices 
Microstruct. {\bf 37}, 306 (2005).

\bibitem{norm} Normalized to unit charge and an effective 
volume in the QW region.
 
\bibitem{Li2009:APL}
P. Li and H. Dery,
Appl. Phys. Lett. {\bf 94}, 192108 (2009);
S.~A. Crooker, E.~S. Garlid, A.~N. Chantis, D.~L. Smith,
K.~S.~M. Redd, Q.~O. Hu, T. Kondo, and C.~J. Palmstr\o m,
Phys. Rev. B {\bf 80}, 041305 (2009).


\bibitem{Ando1998:APL} 
S. Hallstein, J.~D. Berger, M. Hilpert, H.~C. Schneider, W.~W. R\"{u}hle,
F. Jahnke, S.~W. Koch, H.~M. Gibbs, G. Khitrova, M. Oestreich, Phys. Rev.
B {\bf 56}, R7076 (1997), fast spin precession of electrons in an applied 
magnetic field could be used for PM with $\omega/2\pi > 40$ GHz.  ; H. Ando, T. Sogawa, and H. Gotoh, Appl. Phys. 
Lett. {\bf 73}, 566 (1998).

\bibitem{Zutic2006:PRL}
I. \v{Z}uti\'c, J. Fabian, and S.~C. Erwin,  Phys. Rev. Lett. {\bf 97}, 
026602 (2006);
equilibrium densities are negligibly small for laser operation.

\bibitem{QR} 
A recombination quadratic in carrier densities was analyzed in 
Ref.~\onlinecite{Gothgen2008:APL}. 

\bibitem{Huang1993:OQE}
J. Huang and L.~W. Casperson, Opt. Quantum Electron. {\bf 25}, 369 (1993).

\bibitem{pure}
These constraints could be relaxed with the generation of
pure spin currents implying  $P_J>1$,  demonstrated optically
and electrically, as discussed in Ref.~1 and M.~J. Stevens, A.~L. Smirl,
R.~D.~R. Bhat, A. Najmaie, J.~E. Sipe, and H.~M. {van Driel}, Phys. Rev. 
Lett. {\bf 90}, 136603 (2003).  

\bibitem{Yariv:1997}
A. Yariv, {\it Optical Electronics in Modern Communications, 5$^{th}$ Edition} 
(Oxford, 1997).

\bibitem{Garzon2008:PRB}
S. Garzon, L. Ye, R.~A. Webb, T.~M. Crawford, M. Covington, and S. Kaka, 
Phys. Rev. B {\bf 78}, 180401(R) (2008).

\end{thebibliography}
\end{document}